\documentclass{cimento}
\usepackage{graphicx}
\setcopyright{CERN on behalf the CMS Collaboration}
\title{Top Quark Properties Measurements in CMS}
\author{E.~Yazgan on behalf of the CMS Collaboration
\thanks{email:efe.yazgan@cern.ch}}
\instlist{\inst Ghent University, CERN PH-UCM Bat 42 2-029 C28810, CH 1211 Geneva 23, Switzerland}

\PACSes{
\PACSit{14.65.Ha}{Top quarks}
}

\begin{document}

\maketitle

\begin{abstract}
Recent top quark properties measurements made with the CMS detector at the LHC are presented. The measurements summarized include spin correlation of top quark pairs, asymmetries, top quark mass, and the underlying event in top quark pair events. The results are compared to the standard model predictions and new physics models. 
\end{abstract}

\section{Introduction}
The top quark, with its high mass, provides unique measurements in collider physics. 
It has a lifetime that is shorter than the hadronization time scale. 
Therefore, its "bare" quark properties are accessible such as spin, charge, and mass. 
The LHC Run I measurements presented in this paper include top quark pair $t\overline{t}$ spin correlation (section 2), asymmetries (section 3), top quark mass (section 4). 
In addition, a first look at the underlying event in $t\overline{t}$ events produced in proton-proton collisions at the LHC with a center of mass energy $\sqrt{s}=$ 13 TeV is presented in section 6.  

\section{Spin Correlation of Top Quark Pairs}
Heavy quark spins are correlated in heavy quark production by QCD. Top quarks decay in a timescale shorter than the spin decorrelation timescale $m_t/\Lambda_{QCD}^2\sim3\times10^{-21}$ s, so that the spin correlation information is propagated to the decay products \cite{bigi2986}. This correlation can be measured with or without reconstructing the $t\overline{t}$ system. The double differential $t\overline{t}$ cross section with respect to the top and top antiquark decay angles is proportional to 
$1+\alpha_+C\cos\theta_{+}\cos\theta_{-}$. 
Here, $\theta_+$ is the angle between the top quark and the positively charged lepton (in the $t\overline{t}$ rest frame) and $\theta_-$ between the top antiquark and the negatively charged lepton. 
The correlation coefficient, C, is proportional to the spin correlation strength ($A$) and the product of the spin analyzing powers of the decay particles, $\alpha_+\alpha_-$. 
The spin analyzing power for each decay product at next-to leading order (NLO) is given in \cite{Bernreuther2004}. Charged leptons have the highest spin analyzing power (99.8\%), therefore measuring the spin correlation is easiest in the di-lepton channel where both the top and the top antiquark in the event decay to lepton-neutrino pairs. The down-type quarks have relatively large spin analyzing power (96.6\%), while the up-type quarks have only 30\%. Moreover, it is difficult to distinguish up- and down-type quarks. Because of these reasons, measuring the spin correlation in the lepton+jets channel is more difficult than in the dilepton channel. 
The spin correlation strength, $A$, depends on the basis selected for spin-quantization, the final state, and is given by the asymmetry of the number of aligned and anti-aligned top and top antiquark spins.
A basis and channel independent quantity, the fraction of $t\overline{t}$ events with the SM prediction of spin correlation $f$ can be defined as 
$f=N_{SM}^{t\overline{t}}/(N_{SM}^{t\overline{t}}+N_{uncor}^{t\overline{t}})$
where $N_{SM}^{t\overline{t}}$ is the number of $t\overline{t}$ events that display SM spin correlation and  $N_{uncor}^{t\overline{t}}$ indicates the number uncorrelated $t\overline{t}$ events.
The spin correlated fraction, $f$ can be used to determine $A_{basis}^{meas}$ using the SM value of A determined for the selected basis and channel, $A_{basis}^{SM}$ with
$A_{basis}^{meas} = A_{basis}^{SM}f$.

The most recent and precise measurements of the spin correlation variables in dilepton \cite{ref:scdilepton} and lepton+jets \cite{ref:scmem} channels using CMS \cite{ref:CMS} data corresponding to an integrated luminosity of 19.5 fb$^{-1}$ are summarized below. 

\begin{figure}
\begin{center}
\includegraphics[width=0.3\textwidth]{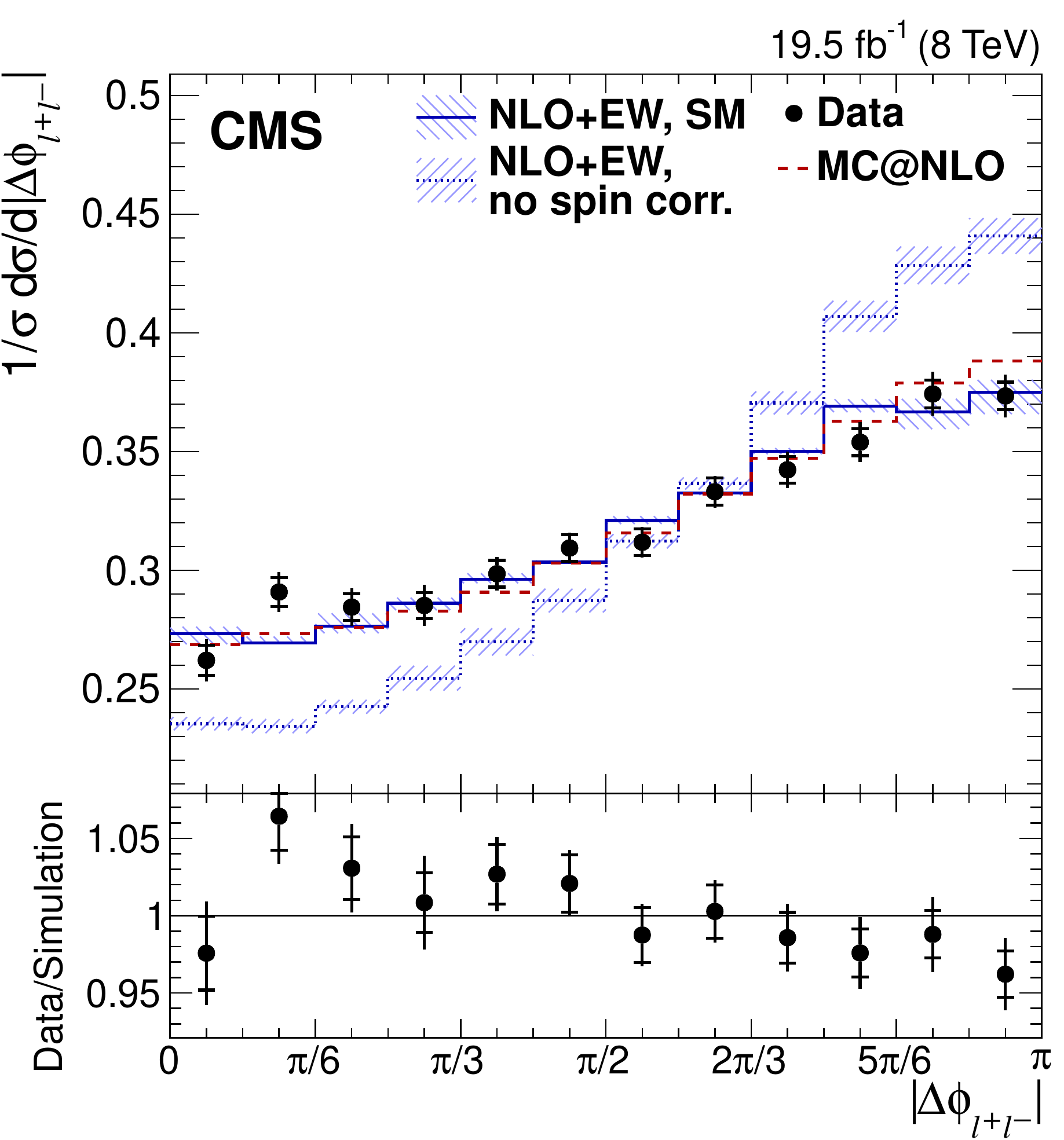}
\includegraphics[width=0.342\textwidth]{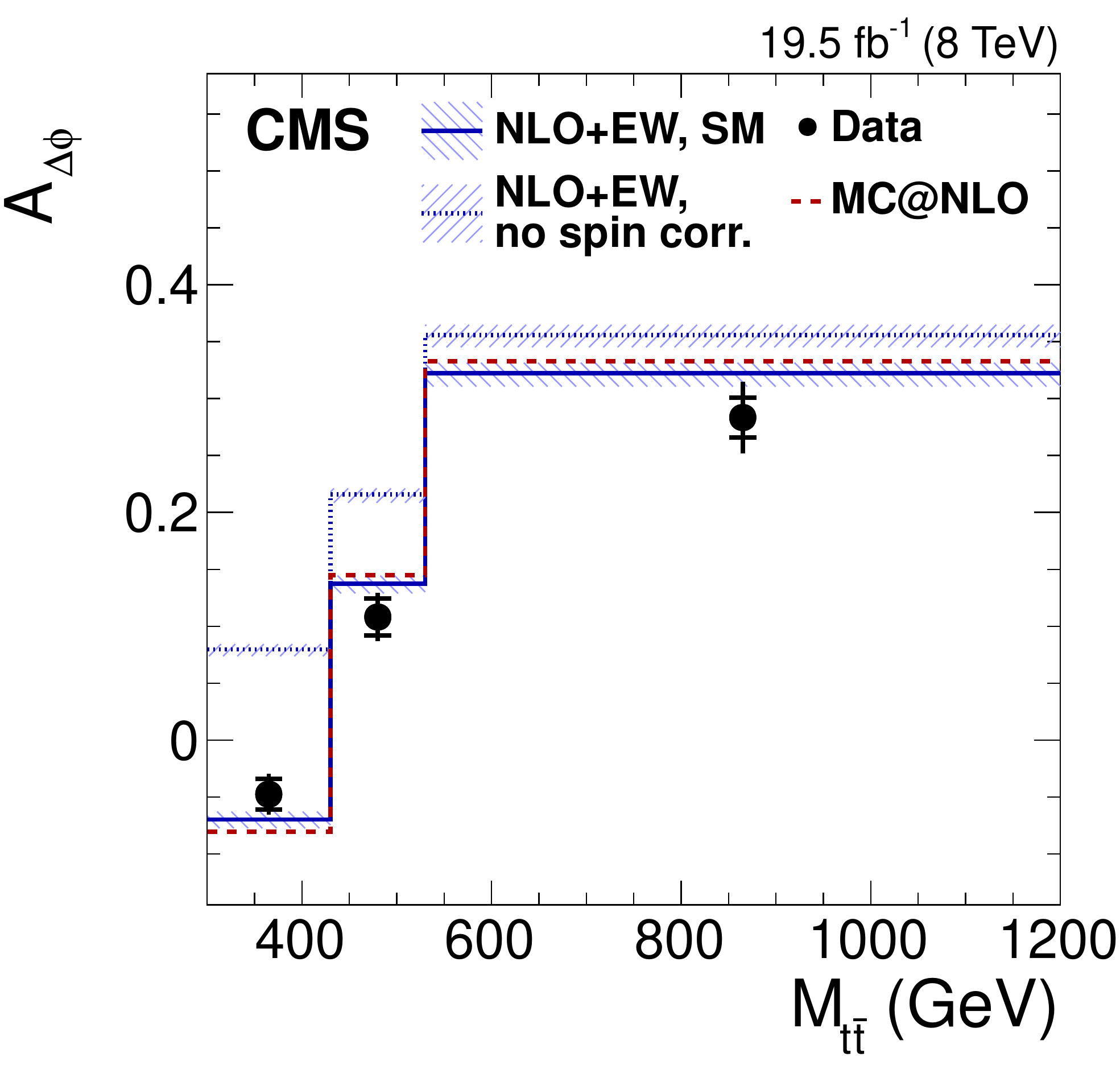}
\includegraphics[width=0.342\textwidth]{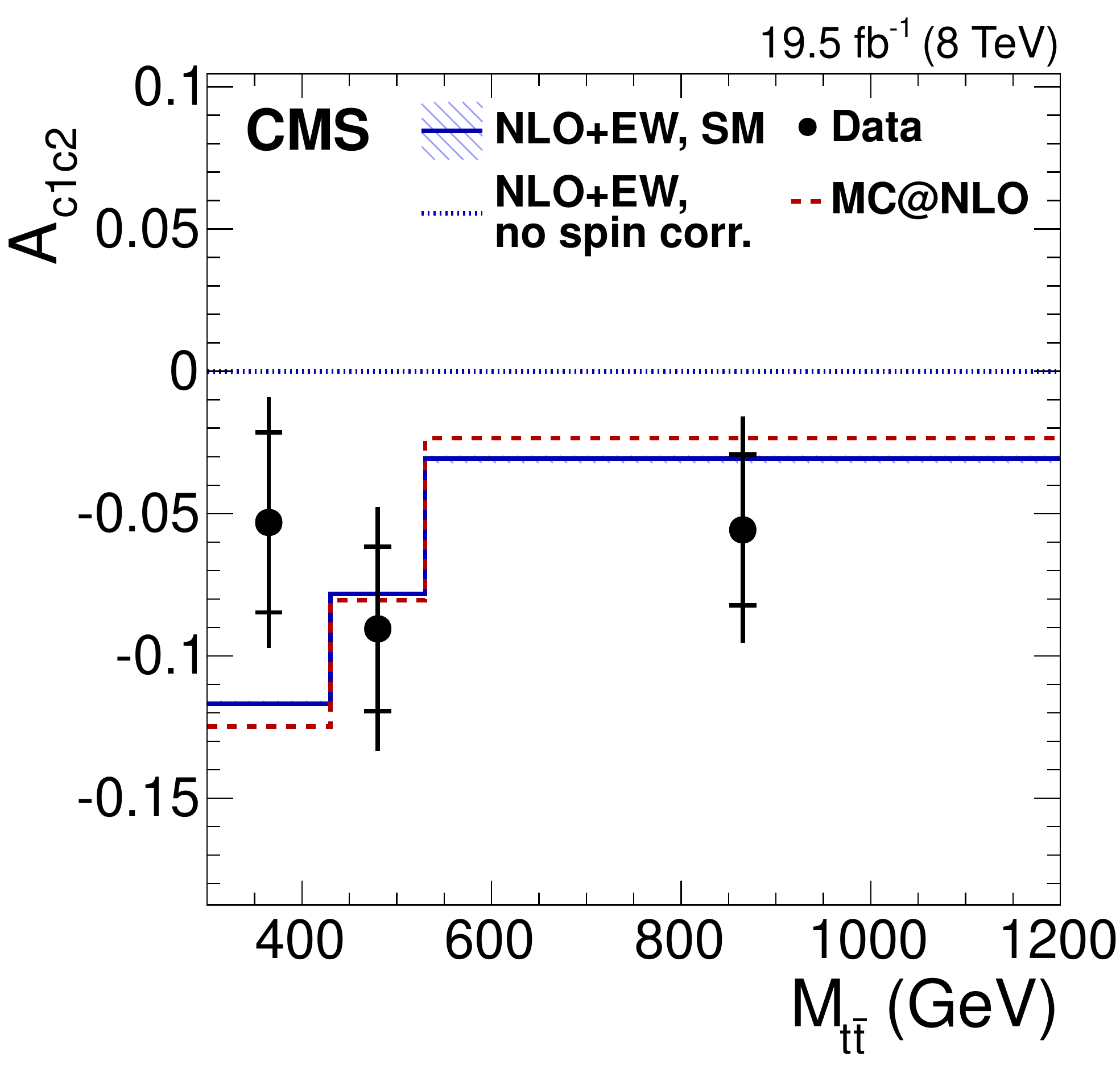}
\end{center}
\caption{Spin correlation measurement in the dilepton channel at $\sqrt{s}=8$ TeV \cite{ref:scdilepton}. Unfolded, normalized differential cross section in bins of $|\Delta\phi_{\ell^+\ell^-}|$ (left). The data over MC@NLO predictions are displayed in the lower panel. 
$A_{|\Delta\phi_{\ell^+\ell^-}|}$ and $A_{c_1c_2}$ in bins of $M_{t\overline{t}}$ (middle and right, respectively).  Comparisons to MC@NLO predictions (dashed histograms), SM NLO+EW predictions (solid histograms) and uncorrelated spins (dotted histograms). The statistical and total uncertainties are shown with the inner and outer vertical bars on the data points. The QCD scale variations are displayed with hatched bands. 
}
\label{tab:dilepton_sc}
\end{figure}

In the dilepton channel, the spin correlation is measured using three different asymmetry variables. These asymmetries are defined through the asymmetries of different  angular variables ($\beta$) constructed from the two final state leptons 
\begin{equation}
A_\beta = \frac{N(\beta>0)-N(\beta<0)}{N(\beta>0)+N(\beta<0)}
\end{equation}
Here,  $\beta$ is taken to be $|\Delta\phi_{\ell^+\ell^-}| - \pi/2$, $\cos\theta_{\ell^+}^*\cos\theta_{\ell^-}^*$, or $\cos\varphi$. 
The asymmetry, $A_{\Delta\phi}$ is used to discriminate between SM and uncorrelated $t\overline{t}$ spins and defined using the opening angle between the two leptons in the lab frame, $\Delta\phi_{\ell^+\ell^-}$. 
The second asymmetry variable $A_{c_1c_2}$ that provides a direct measurement of the spin correlation coefficient $C_{hel}=-4A_{c_1c_2}$  is defined using the helicity angles, $c_1=\cos\theta_{\ell^+}^*$ and $c_2=\cos\theta_{\ell^-}^*$. The helicity angle, $\theta_\ell^*$, is defined to be the angle of the lepton  and its parent quark (or antiquark) momentum direction in the $t\overline{t}$ center-of-mass frame, i.e. the helicity frame. 
The third asymmetry variable, $A_{\cos\varphi}$, provides a direct measurement of the spin correlation coefficient $D=-2A_{\cos\varphi}$ where $\varphi$ is the angle between the two lepton momenta in their respective parent top and antiquark's rest frames. 
These three asymmetries are measured inclusively and differentially with respect to invariant mass (M$_{t\overline{t}}$), rapidity 
($|y_{t\overline{t}}|$), and the transverse momentum ($p_T^{t\overline{t}}$) of the $t\overline{t}$ system. The data is corrected to the parton level using an unfolding procedure.  
The unfolded and normalized inclusive cross section vs $|\Delta\phi_{\ell^+\ell^-}|$, $A_{\Delta\phi}$ vs M$_{t\overline{t}}$, and $A_{c_1c_2}$ vs M$_{t\overline{t}}$ are displayed in Figure \ref{tab:dilepton_sc}.
The data is found to agree well with SM predictions with the dominant systematic uncertainty being the top quark $p_T$ modeling. 
This uncertainty becomes smaller in the double-unfolded distribution, $A_{\Delta\phi}$ vs $M_{t\overline{t}}$. From the measured asymmetries, the spin correlated fractions (f) are obtained using the QCD NLO predictions including EWK corrections. The most precise f value obtained (from $A_{\Delta\phi}$ vs $M_{t\overline{t}}$) is $1.12\pm0.06(stat)\pm0.08(syst)^{+0.08}_{-0.11}(theor)$ consistent with the SM prediction, i.e. f=1. 

In the dilepton channel, the measured $t\overline{t}$ spin correlation variables are also used to search for hypothetical anomalous, flavor-conserving interaction between the top quark and the gluon. This is done in a model independent fashion using top quark chromo-magnetic and -electric moments assuming a particle exchange with a mass  larger than the mass of the top quark. No evidence for an anomalous $t\overline{t}g$ coupling is observed and it is found that the real part of the chromo-magnetic dipole moment, Re($\hat\mu_t$) can only assume values between -0.053 and 0.026, and the imaginary part of the chromo-electric dipole moment, Im($\hat d_t$) can only have values between -0.068 and 0.067 to be consistent with observations. 

\begin{figure}
\label{fig:spincorr}
\begin{center}
\includegraphics[width=0.35\textwidth]{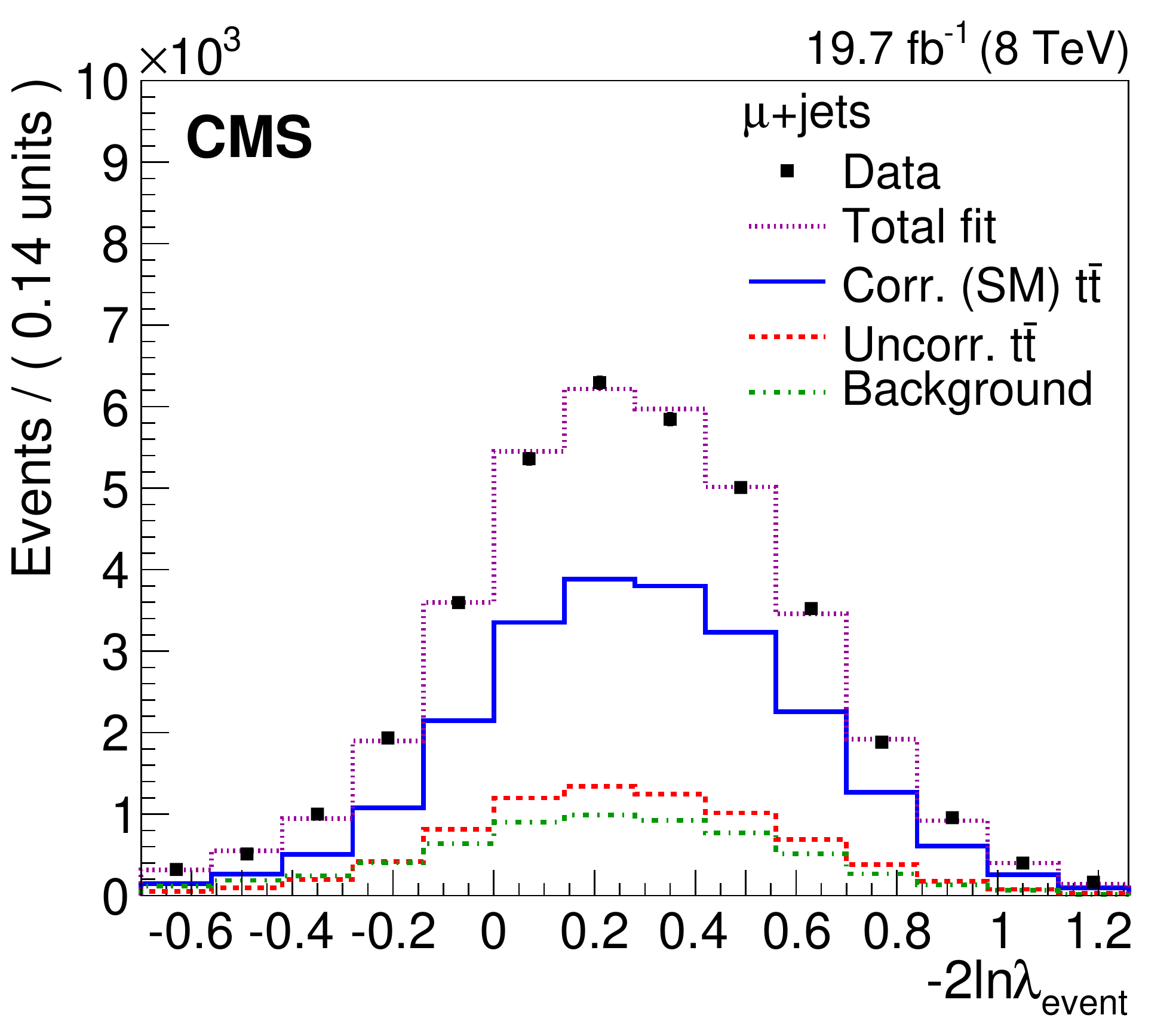}
\includegraphics[width=0.35\textwidth]{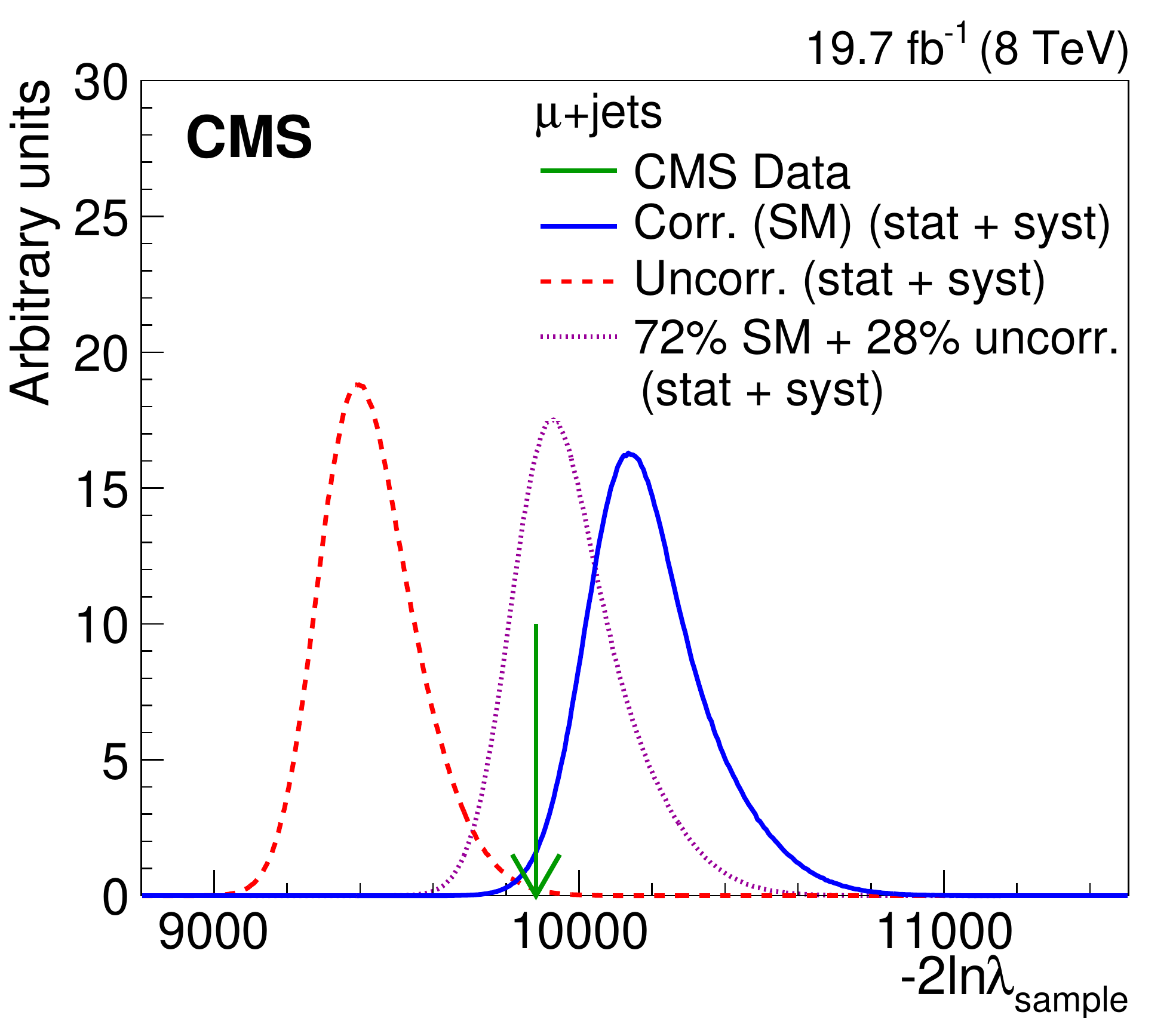}
\end{center}
\caption{Spin correlation measurement in the lepton+jets channel at $\sqrt{s}=8$ TeV \cite{ref:scmem}. Template fit to the data (left). Data are represented by squares. The dotted curve shows the overall result. The contributions of the SM template, spin-uncorrelated signal, and the background are displayed by the solid curve, the dashed curve, and the dash-dot curve, respectively. Sample likelihood distributions in simulation calculated at the data set size (right). The arrow shows the sample likelihood value obtained from the data.  The dotted curve displays the mixture of SM and spin-correlated events as determined from the template fit.} 
\end{figure}

Spin correlation measurement is more challenging in the lepton+jets channel due to the smaller spin analyzing power of quarks, the difficulty in distinguishing jets from up and down type quarks, and the lower resolution of jets compared to leptons. Because of these reasons, a multi-varied method, the matrix element method, is employed to make the measurement. In this method, the compatibility of an event with the SM spin correlation or uncorrelated spin hypotheses ($H_{SM}$ and $H_{uncorr}$ respectively) are calculated using the following formula
\begin{equation}
P(x_i | H) = \frac{1}{\sigma_{obs}(H)}\int f_{PDF}(q_1)f_{PDF}(q_2)dq_1dq_2\frac{(2\pi)^4|M(y,H)|^2}{q_1q_2s}W(x_i,y)d\Phi_6
\end{equation}
Here, $q_1$ and $q_2$ are the parton momentum fractions, $f_{PDF}(q_1)$ and $f_{PDF}(q_2)$ are the corresponding parton distribution functions (PDFs), s represents the square of the center-of-mass energy of the colliding protons, and $d\Phi_6$ is the six-dimensional phase space element. 
The transfer function denoted by $W(x_i,y)$  maps the reconstructed kinematic properties $x$ to parton level properties $y$ (in three different detector regions in this particular measurement). The leading order (LO) matrix element (ME) is denoted by $M(y,H)$ in which $H$ is the SM or uncorrelated spins hypothesis. Using the MadWeight code \cite{madweight2010}, likelihoods for the two hypotheses, $P(H_{uncorr})$ and $P(H_{SM})$, are calculated for each event and the negative log-likelihood ratio, -2$ln\lambda_{event}=-2ln[P(H_{uncorr})/P(H_{SM})]$, is used as the discriminating variable.  
MadWeight partially corrects for the effect of the initial-state radiation (ISR) using the overall partonic transverse momentum of the $t\overline{t}$ system. A kinematic fitter is used to select the four jets from the LO $t\overline{t}$ process as input to the LO ME calculation. 
A fit to the SM and spin-uncorrelated likelihood ratio templates constructed from simulated NLO events, the spin correlated fraction, $f$, and the background fraction are extracted. Post-fit distributions are displayed in Figure \ref{fig:spincorr}. The resulting spin correlated fraction is, $f=0.72\pm0.08(stat)^{+0.15}_{-0.13}(syst)$. In this measurement, the dominant systematic uncertainties are the jet energy scale, QCD scale, and the top quark mass. This measurement represents the most precise result in the muon+jets channel and has a precision close to the most precise result in the dilepton channel.  The $t\overline{t}$ spin correlation hypothesis as predicted by the SM is tested combining event likelihood ratios to construct the sample likelihood ratio for SM and spin-uncorrelated hypotheses.  It is found that the data agree with the SM model within 2.2 standard deviations and agree with the uncorrelated hypothesis within 2.9 standard deviations. Hypothesis testing and template fit results are found to be consistent as shown in Figure \ref{fig:spincorr}. 

\section{Top Quark Pair Asymmetries}
\begin{figure}
\label{fig:ac}
\begin{center}
\includegraphics[width=0.3\textwidth]{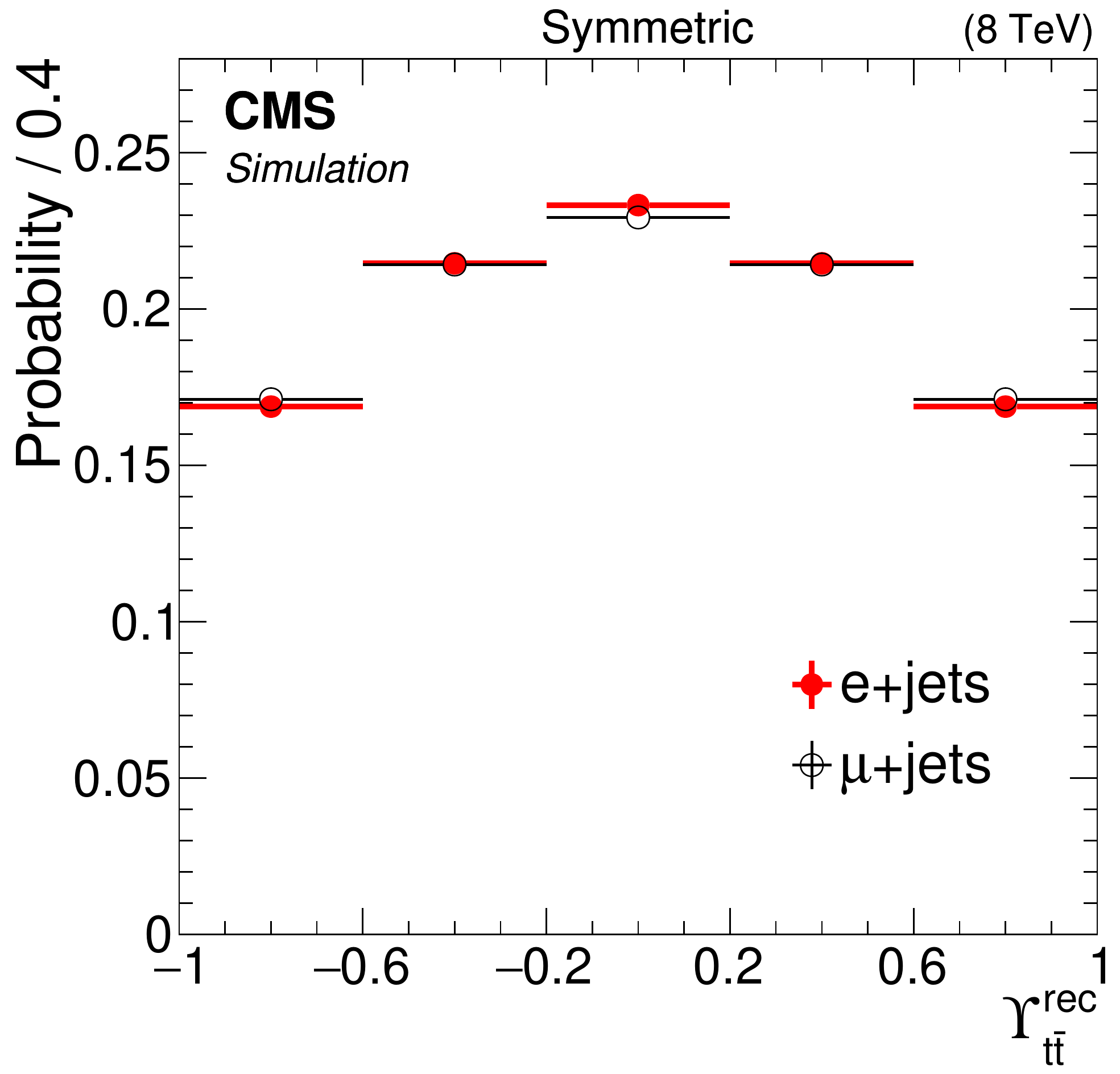}
\includegraphics[width=0.3\textwidth]{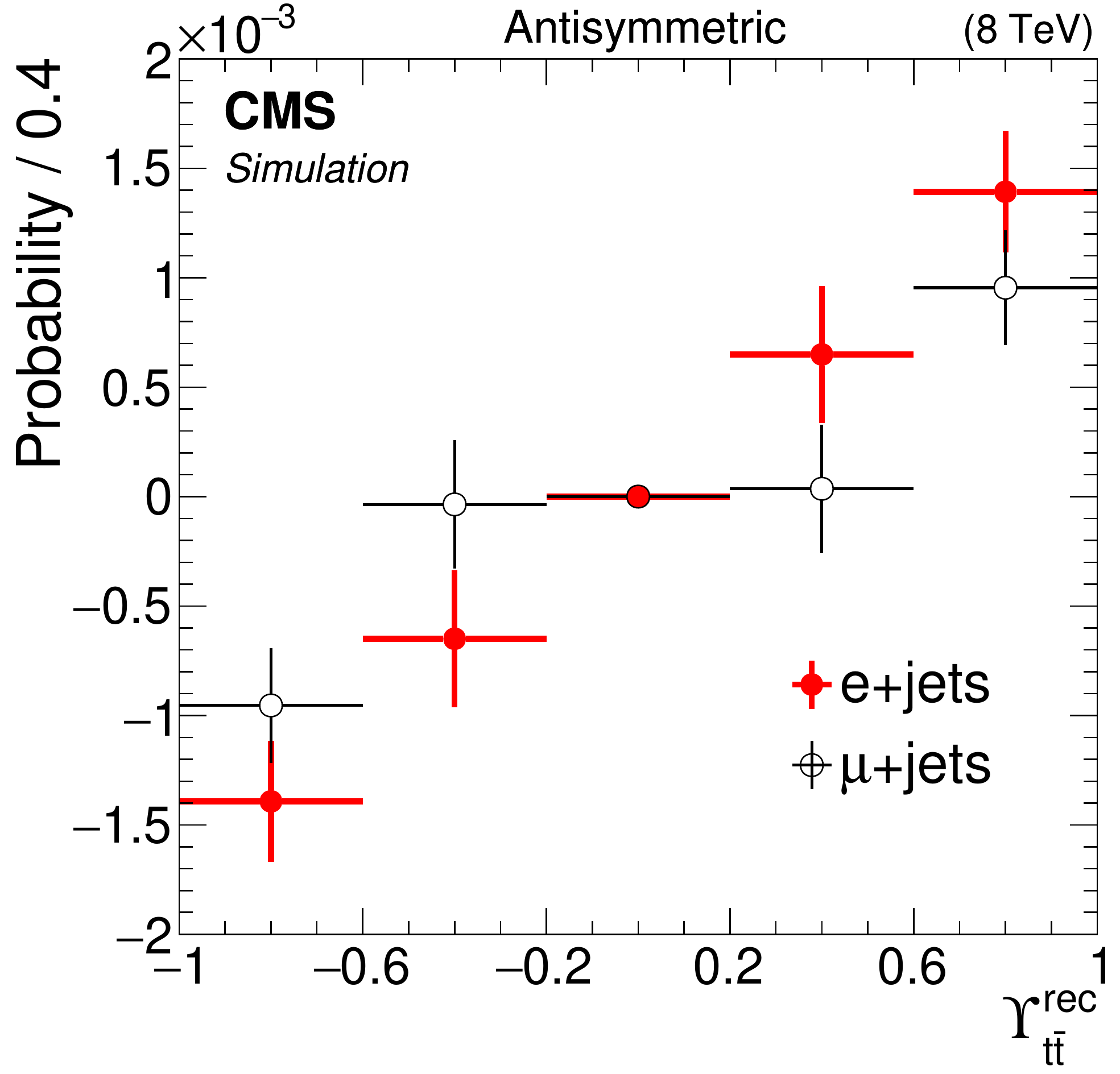}
\includegraphics[width=0.3\textwidth]{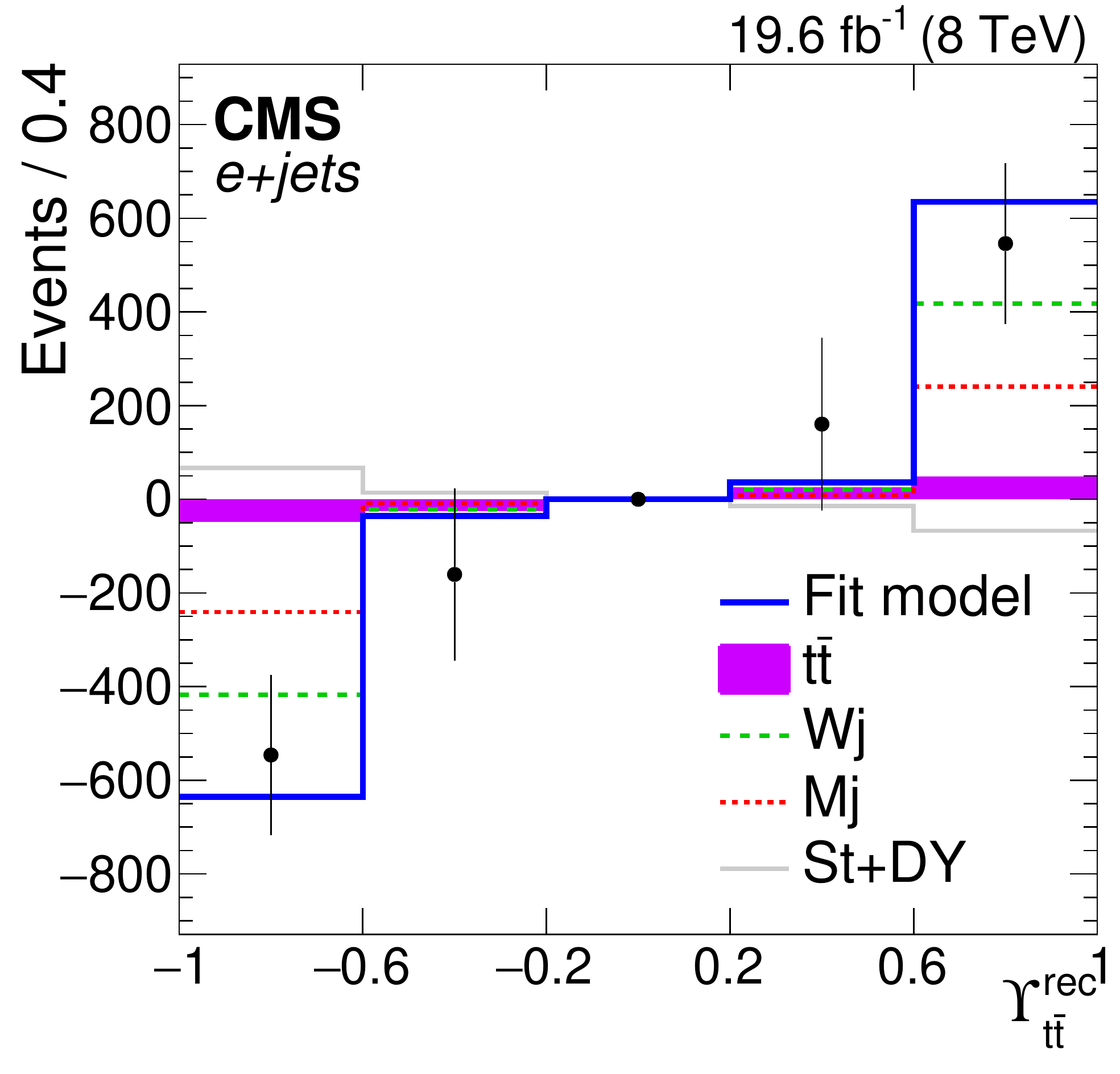}
\end{center}
\caption{Charge asymmetry measurement in the lepton+jets channel at $\sqrt{s}=8$ TeV \cite{ref:burt}. The symmetric component of $\Upsilon^{rec}_{t\overline{t}}$ in the e+jets and $\mu$+jets channels (left). The antisymmetric component of $\Upsilon^{rec}_{t\overline{t}}$ in the e+jets and $\mu$+jets channels (right).}
\end{figure}

At NLO QCD, the interference between tree-level and box diagrams and the interference between ISR and final-state radiation (FSR) cause a small charge asymmetry that results in more widely distributed top quarks compared to top antiquarks at the LHC. There are significant contributions from QCD-electroweak interference terms \cite{ref:hollikpagani2011} and only small contributions from quark-gluon scattering \cite{kuhnrodrigo1999}. Using CMS data $\sqrt{s}=$8 TeV, charge asymmetry is measured both by correcting the data for detector and acceptance effects \cite{ref:thorsten} and a template fit method \cite{ref:burt}. In the template fit method, a new variable that changes sign under the exchange of $t$ and $\overline{t}$ is defined as $\Upsilon_{t\overline{t}}=\tanh\Delta|y|_{t\overline{t}}$, where  $\Delta|y|_{t\overline{t}}$ is the difference of absolute values of top and top antiquark rapidities, $|y_t|-|y_{\overline{t}}|$. The probability density, $\rho(\Upsilon)$, could be expressed in symmetric ($\rho^+$) and antisymmetric ($\rho^-$) components with $\rho^{\pm}(\Upsilon)=[\rho(\Upsilon)\pm\rho(-\Upsilon)]/2$. POWHEG \cite{ref:powheg} event generator with CT10 PDF set \cite{ref:ct10} is used to construct the two components of $\rho(\Upsilon)$ that provides the base model charge asymmetry, $\hat{A}_c^\Upsilon$. Then, a generalized model can be defined from a linear combination of the symmetric and antisymmetric base model components $\rho(\alpha)=\rho^++\alpha\rho^-$. With this formulation, the asymmetry in data is equal to the base model with the free parameter $\alpha$, i.e. $A_c^\Upsilon(\alpha)=\alpha\hat{A}_c^\Upsilon$. 
A template fit to the antisymmetric component is made to extract $\alpha$. 
The templates for the symmetric and antisymmetric components of the base model for the $e+jets$ and $\mu+jets$ channels, as well as the $e+jets$ data projected along $\Upsilon^{rec}_{t\overline{t}}$ with the fitted model are displayed in Figure \ref{fig:ac}. Using both channels, the combined charged asymmetry is measured to be $A_c^y=[0.33\pm0.26(stat)\pm0.33(syst)]\%$. This result is the most precise charge asymmetry measurement. It is consistent with the NLO QCD predictions but does not rule out the alternative models considered, i.e. 200 GeV and 2 TeV axigluons and Z' bosons. While the charge asymmetry measurement gives the most precise inclusive result, the unfolded measurement also allows differential measurements with respect to different kinematic variables such as the rapidity, transverse momentum and invariant mass of the $t\overline{t}$ system both in full phase space and in a reduced fiducial phase space \cite{ref:thorsten}. All measurements are found to be consistent within two standard deviations with a vanishing asymmetry and also with the SM predictions.  The differential measurements are also compared to a model with an effective axial-vector coupling of the gluon \cite{ref:emidio1,ref:emidio2} with new physics scales of 1.5 and 2 TeV. It is found that for $m_{t\overline{t}}>450$ GeV, the measured charge asymmetry is about 2 standard deviations from the model with a scale of 1.5 TeV, while, measurements are still consistent within $\sim1$ standard deviations with the model with a 2 TeV new physics scale. 

\section{Top Quark Mass}
Recent measurements of the top quark mass using data taken at $\sqrt{s}=8$ TeV in lepton+jets, all-jets, and dilepton channels by CMS yielded the most precise measurements in each. These measurements combined with the published CMS measurements at $\sqrt{s}$=7 TeV yields
a top quark mass of $m_t=172.44\pm0.13(stat)\pm0.47(syst)$ GeV with a precision of 0.3\% representing the most precise top quark mass measurement to date \cite{ref:cmsmass}. The dominant uncertainties are flavor-dependent jet energy corrections and b-jet modeling. 
The dependence of top quark mass on event kinematics is also studied in the lepton+jets channel. This is important to determine perturbative and non-perturbative effects that have different kinematic dependences in different parts of the phase space. Variables sensitive to color connection, ISR/FSR, and b-quark kinematics are measured and no indication of a kinematic bias has been observed. The data is found to be consistent with theory predictions calculated using different MC generator and parton shower codes. 

\begin{figure}
\begin{center}
\includegraphics[width=0.35\textwidth]{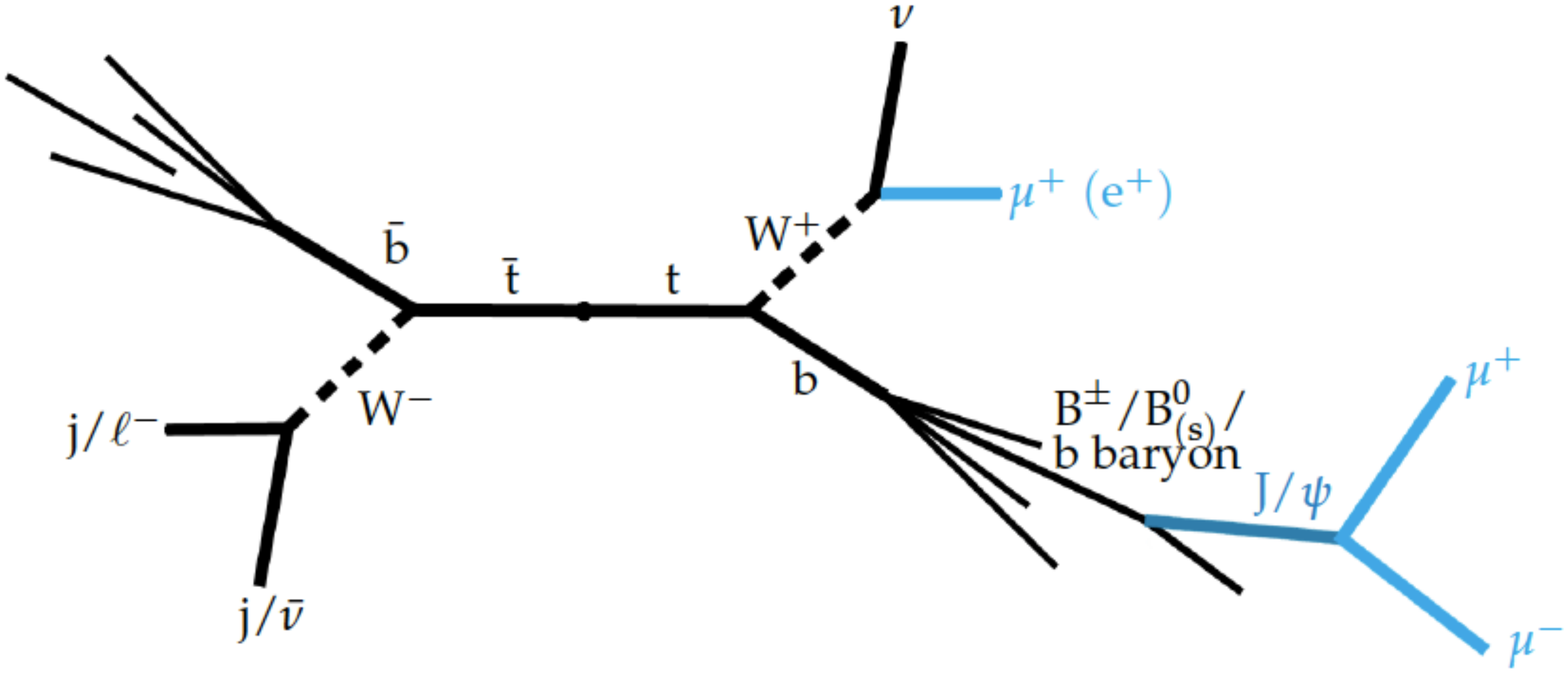}
\includegraphics[width=0.3\textwidth]{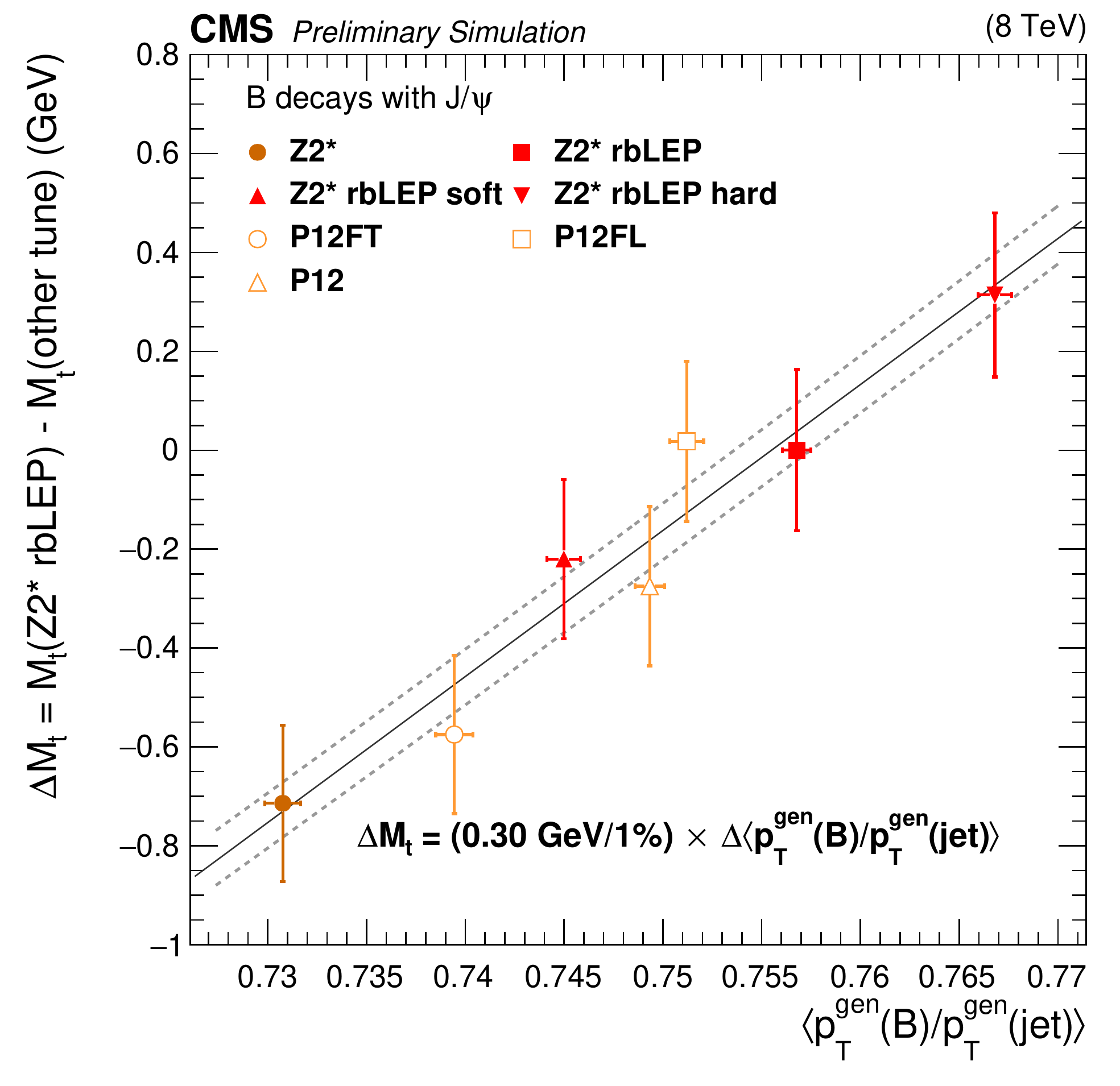}
\includegraphics[width=0.3\textwidth]{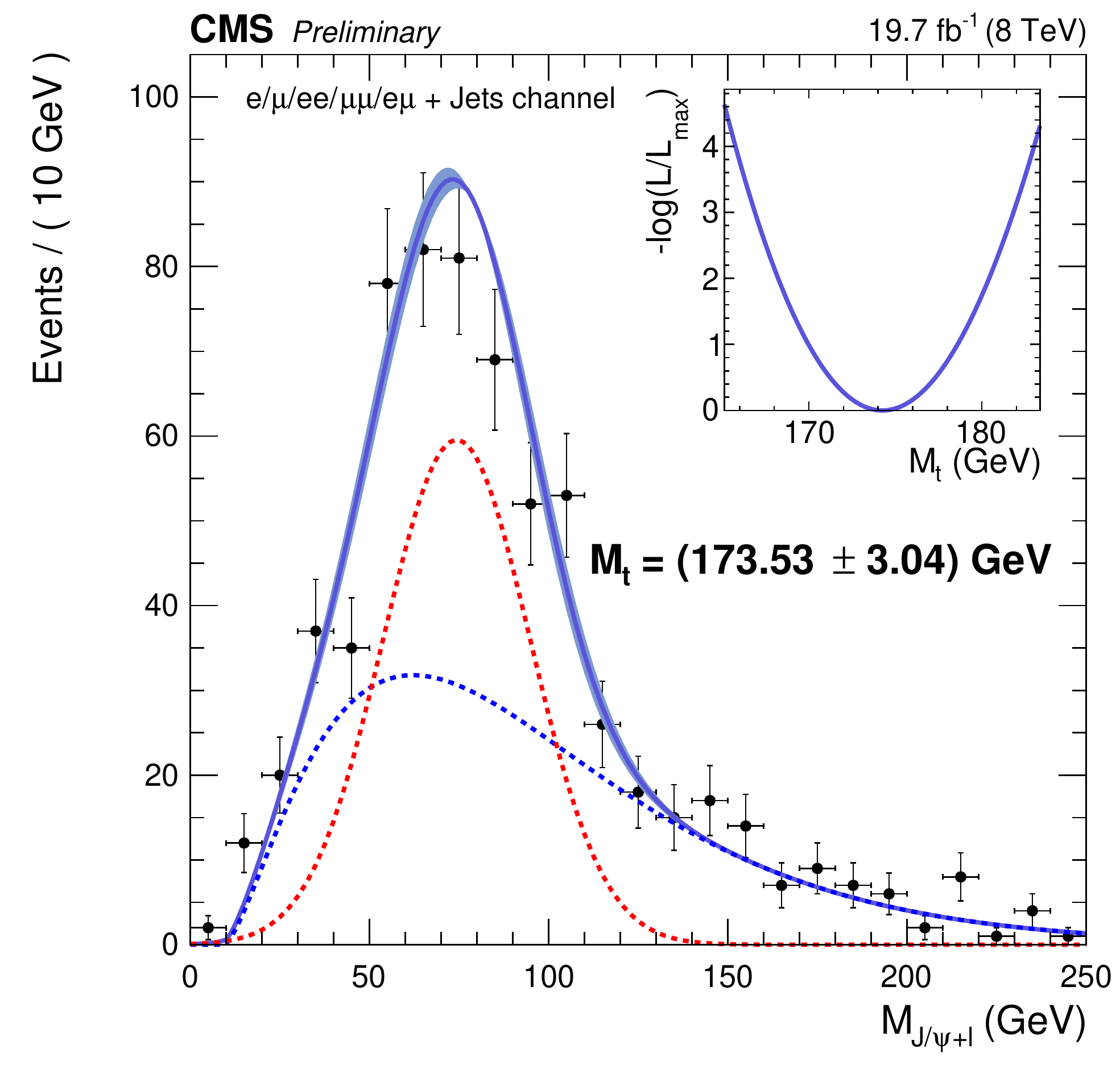}
\end{center}
\label{fig:jpsi}
\caption{Top quark mass measurement using $J/\psi+\mu$ mass distribution from $t\overline{t}$ events at $\sqrt{s}=8$ TeV \cite{ref:jpsi}. Exclusive $J/\psi$ in a $t\overline{t}$ process (left). Average fragmentation vs the extracted top quark mass fitted with a line (middle). Fit to the $J/\psi$+lepton signal invariant mass (right). The log likelihood scale as a function of the top quark mass is shown in the inset.}
\end{figure}

The top quark mass measurements from direct reconstruction are dominated by uncertainties in jet energy scales and soft QCD modeling such as b-quark hadronization. To minimize experimental uncertainties, an alternative top quark mass measurement method was proposed in 1992 \cite{ref:jpsi1,ref:jpsi2}. In this method, the correlation between the 3-prong leptonic mass and the top quark mass is utilized. For the first time, this method is applied by CMS \cite{ref:jpsi} by measuring the top quark mass in the exclusive decay channel $t\rightarrow(W\rightarrow\ell\nu)(b\rightarrow J/\psi+X\rightarrow\mu^+\mu^-+X)$ (see Figure~\ref{fig:jpsi}). Both $t\overline{t}$ and single top processes are taken as signal. The $J/\psi$ mass is reconstructed from the selected two non-isolated muons from the same jet with $p_T>4$ GeV. The $J/\psi$ mass is required to be in the 3-3.2 GeV window.  The wrong lepton pairings in the analysis constitutes 51\% of the events however these wrong pairings still have some useful correlation to the top quark mass. Therefore, in addition to good pairings, the wrong ones are also utilized. Finally, the reconstructed invariant  $J/\psi+\mu$  mass distribution is fit to an analytic function to extract the top quark mass. 
The MC simulations of the $t\overline{t}$ processes utilized in this measurement are made assuming the Z2* tune \cite{ref:z2}. 
Using the transverse momentum of the B hadron $p_T^{gen}(B)$ relative to the jet the hadron belongs to at the generator level $p_T^{gen}(jet)$, $M_{J/\psi+\ell}(Z2*, m_t=172.5~GeV)$ is reweighted for different tunes. 
The dependence of the extracted top quark mass on the average fragmentation calculated with the average transverse momentum the B hadron and the jet $<p_T^{gen}(B)/p_T^{gen}(jet)>$ is shown in Figure~\ref{fig:jpsi}. The default tune is found to be softer than all other tunes. The P12 and other Z2* tune families are consistent with each other within their statistical uncertainties. The Z2*rbLEP tune which includes the B hadron, Z boson momenta and mass measurements at LEP is chosen as the baseline tune. This yields a shift, $m_t(Z2*rbLEP)-m_t(Z2*)=-0.71$ GeV. The resulting top quark mass is $m_t=173.53\pm3.0(stat)\pm0.9(syst)$ GeV extracted from the fit shown in Figure~\ref{fig:jpsi}. Due to the small branching ratio of the  $t\rightarrow(W\rightarrow\ell\nu)(b\rightarrow J/\psi+X\rightarrow\mu^+\mu^-+X)$ process, the measurement is statistically limited. This measurement method does not rely on jets, therefore the experimental systematic uncertainties are kept at a minimum. 
The most dominant systematic uncertainties are the modeling of the top quark transverse momentum, b-jet fragmentation, and the MC generator that could be improved with the new versions of generators. This is the first experimental result using this method. With the next LHC runs and updated simulation codes, it could yield a measurement close in precision to the direct mass measurements.  

The quark mass values and the strong coupling constant, $\alpha_s$, are the free parameters of the QCD Lagrangian. 
Fixing $\alpha_s$ and the PDF in a $t\overline{t}$ cross section calculation leaves the top quark mass as the only free parameter which could be extracted by comparing the results of the calculation to the inclusive $t\overline{t}$ cross-section ($\sigma_{t\overline{t}}$) measurements. The top quark pole mass vs $\sigma_{t\overline{t}}$ at $\sqrt{s}=7$ and 8 TeV are used. The likelihood for the predicted dependence is determined using the TOP++ \cite{ref:czakon} program providing a next-to-next-to leading order (NNLO)  calculation employing NNPDF3.0 PDF set \cite{ref:nnpdf}. 
The product of the likelihoods for the NNLO prediction and the experimental result is used to extract the top quark pole mass by maximizing with respect to top quark mass and inclusive cross section simultaneously. The mass extraction is done using three different PDF sets separately, namely, NNPDF3.0, MMHT2014 \cite{ref:mmht}, and CT14 \cite{ref:ct14}. It is found that the top quark pole mass values are consistent using the three PDF sets. The measured values at $\sqrt{s}=7$ and 8 TeV are combined for each PDF set. The most precise top quark pole mass, $m_t^{pole}=173.8^{+1.7}_{-1.8}$ GeV is obtained by the PDF set, NNPDF3.0. 

\section{A First Look at the Underlying Event in Top Quark Pair Events at $\sqrt{s}=13$ TeV}
\begin{figure}
\begin{center}
\includegraphics[width=0.33\textwidth]{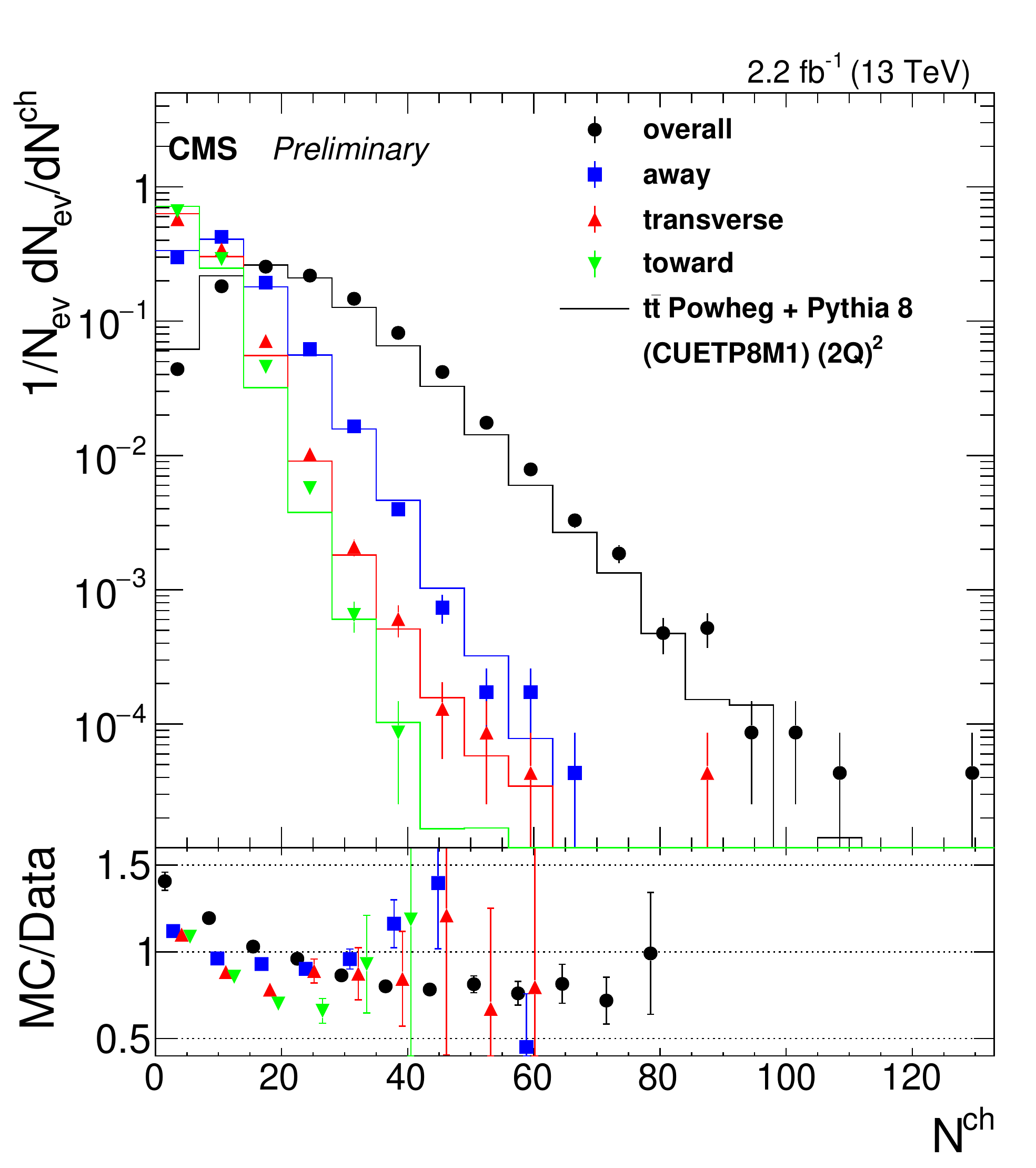}
\includegraphics[width=0.33\textwidth]{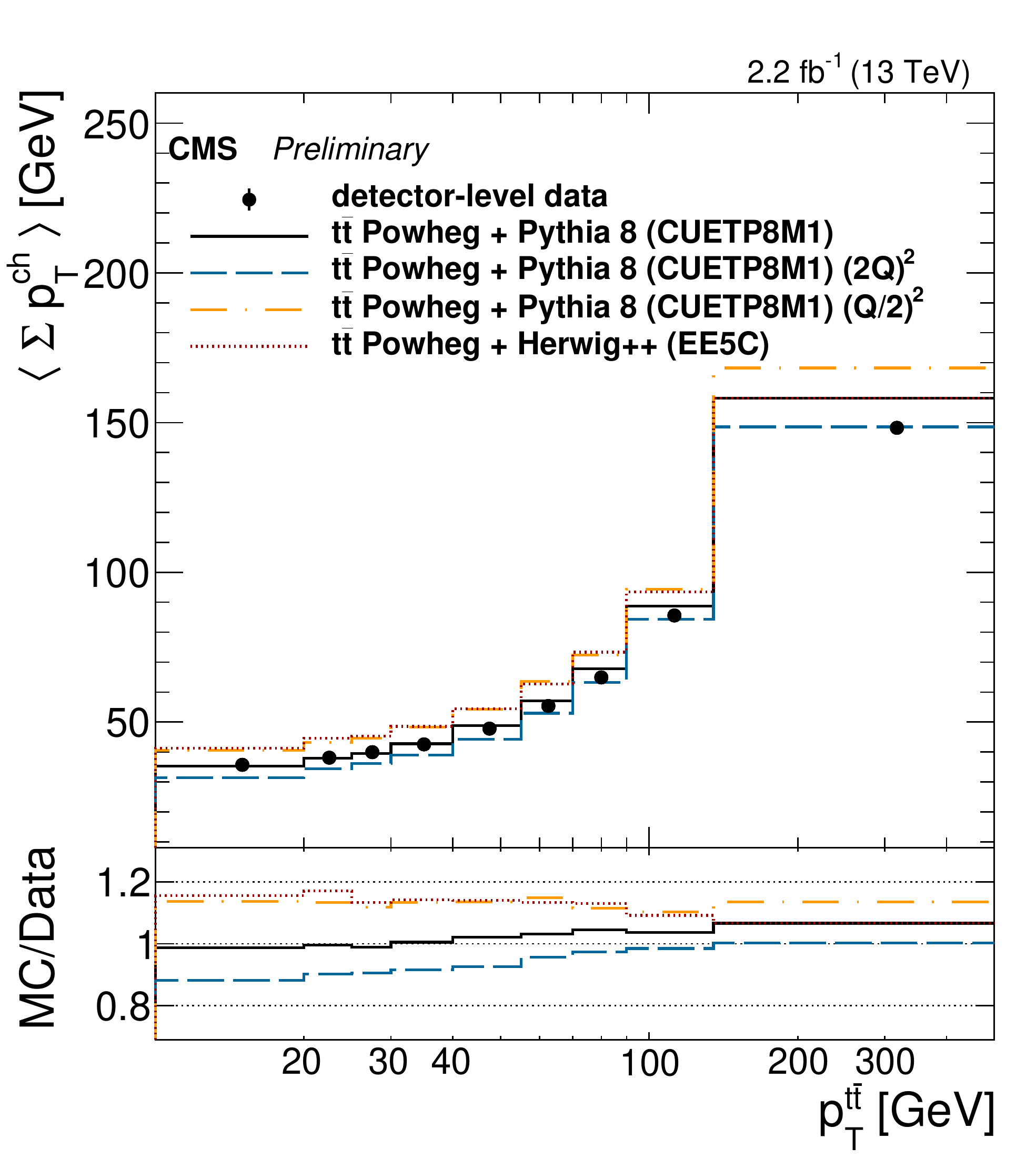}
\end{center}
\caption{Underlying event in $t\overline{t}$ events in the lepton+jets channel at $\sqrt{s}=13$ TeV \cite{ref:ue}. The charged particle candidate multiplicity distributions for different regions with respect to the $t\overline{t}$ system direction (left). The theory predictions assume twice the nominal QCD scale uncertainty. The data is represented by point at the detector level and the theory predictions by lines. Sum of charged $p_T$ vs $p_T^{t\overline{t}}$ for the overall sample (right).}
\label{fig:ue}
\end{figure}

To test and improve modeling of $t\overline{t}$ event modeling, the underlying event properties in  $t\overline{t}$ events at $\sqrt{s}=13$ TeV in the $\mu+jets$ channel are studied \cite{ref:ue}. This is done by measuring the charged particle activity through the number of charged particles ($N^{ch}$), the scalar sum of the transverse momenta of the charged particles ($\Sigma p_T$), and the average transverse momentum per charged particle ($\overline{p_T}$). These variables are measured in different regions defined relative to the $t\overline{t}$ system's direction. Figure~\ref{fig:ue} displays the $N^{ch}$ distributions for the different regions and the overall sample when the CUETP8M1  tune \cite{ref:cuet} is used and with the QCD scale twice the nominal is assumed. The figure also displays 
$<\Sigma p_T>$ vs the transverse momentum  of the $t\overline{t}$ system measured at the detector level compared to predictions from POWHEG+Pythia8 \cite{ref:pythia8} with the CUETP8M1 tune with the nominal QCD scale ($Q^2$) as well as varied scales of $(2Q)^2$, and $(Q/2)^2$. The data is also compared to the predictions from Powheg+Herwig++ \cite{ref:herwig} with the EE5C tune. At this "first look" with the measurements made at the detector level, we conclude that it is not necessary to have separate, dedicated heavy-quark underlying event tunes. 

\section{Conclusions}
Measurements of top quark properties at CMS provide thorough tests of the SM. Selected top quark properties and mass measurements from LHC Run I are presented along with the first underlying event measurement at $\sqrt{s}=13$ TeV at the LHC.






\end{document}